\documentclass[letterpaper, 12pt]{article}
\usepackage[doublespacing]{setspace}
\usepackage{setspace}
\usepackage[utf8]{inputenc}
\usepackage[letterpaper, total={6.8in, 8.3in}]{geometry}
\usepackage{float}
\usepackage{alltt}
\usepackage{color}
\usepackage{graphicx}
\usepackage{epsfig}
\usepackage{lscape}
\usepackage{amsmath,amsfonts,amssymb}
\usepackage{array}
\usepackage{url}
\usepackage{soul}
\usepackage[dvipsnames]{xcolor}
\usepackage{hyperref}
\usepackage{subcaption}
\usepackage{arydshln}
\usepackage{rotating}
\usepackage{floatflt}
\usepackage{pdflscape}
\usepackage{etoolbox}
\usepackage{csquotes}
\usepackage{multicol}
\usepackage{algorithm2e}
\usepackage{booktabs}
\SetKwBlock{phaseone}{begin}{end}
\SetKw{kwPo}{Phase one}
\SetKw{kwPt}{Phase two}
\SetKw{kwPth}{Phase three}
\usepackage[section]{placeins}

\newcommand{\indep}{\perp\!\!\!\!\perp}

\usepackage{tabularx}
\usepackage{caption}
\newcolumntype{Y}{>{\raggedleft\arraybackslash}X}
\usepackage[style=authoryear,maxcitenames=1]{biblatex}
\usepackage{tikz}
\usetikzlibrary{bayesnet}
\usetikzlibrary{shapes,decorations,arrows,calc,arrows.meta,fit,positioning}
\tikzset{
	-latex,auto
  el/.style = {inner sep=3pt, align=left, sloped},
  point/.style = {circle, 
                  draw = black, 
                  inner sep = 0.04cm,
                  fill = black,
                  node contents={}},
  square/.style={rectangle, 
                 draw = black, 
                 inner sep = 0.08cm,
                 node distance = 0.8cm and 0.8cm,
                 fill=black,  
                 node contents={}},
   selection/.style={ellipse, 
                 draw = black, 
                 inner sep = 0.08cm,
                 node distance = 0.8cm and 0.8cm,
                 node contents={}},
      s-node/.style={double distance=0.5mm, circle,draw,inner sep = 0.06cm},
  bidirected/.style={Latex-Latex,dashed}
}

\title{Proximal Causal Inference for Time Series from N-of-1 Mobile Device Data}

\author{} 

\author{Fanyu Cui\thanks{Department of Biostatistics, Columbia University. Email: fc2754@cumc.columbia.edu} \and
Charlotte Fowler\thanks{Mathematical Sciences Department, Worcester Polytechnic Institute} \and
Xiaoxuan Cai \thanks{Department of Statistics, The Ohio State University} \and
Jukka-Pekka Onnela \thanks{Department of Biostatistics, Harvard University} \and
Justin T Baker \thanks{McLean Hospital} \and
Linda Valeri \thanks{Department of Biostatistics, Columbia University. Email: lv2424@cumc.columbia.edu}\thanks{Department of Epidemiology, Harvard University}}

\begin{document}
\renewcommand{\arraystretch}{0.85}
\maketitle

\begin{center}
    \textbf{Abstract}
\end{center} 

Unmeasured confounding challenges causal inference in intensive longitudinal studies, potentially biasing treatment effect estimates. The proximal causal inference (PCI) framework offers a promising approach to identification using proxies or negative control variables in the presence of hidden confounding bias. While prior literature considers the joint effect of time-varying treatments, our work extends the framework to a time series setting to estimate the contemporaneous or lagged effect of time-varying treatments. We demonstrate that under traditional PCI assumptions, we can recover unbiased effect estimation in the presence of unmeasured confounding by leveraging the intensive longitudinal nature of time series data. Specifically, we develop identifiability conditions and show that past and future observations can be employed as natural proxies for unmeasured confounders. We further develop the bridge function required for valid PCI estimation along with asymptotic variance. Simulation studies illustrate our method's validity and robustness to violation of certain PCI identification assumptions. We demonstrate the potential for studying socio-environmental exposure health effects by applying our method to a smartphone study of bipolar patients. Our work contributes to the growing literature on PCI and provides a powerful tool for analyzing longitudinal data, including in observational mobile health research prone to confounding bias.

\section{Introduction}

Unmeasured confounding presents a major challenge for causal inference in intensive longitudinal studies, where data are continuously collected over time. Conventional causal inference methods can yield biased estimates of treatment effects when some confounders are omitted. The proximal causal inference (PCI) framework has recently emerged as a promising solution to this problem, providing identification strategies that leverage proxy variables to adjust for hidden confounding bias (\cite{tchetgen2024}, \cite{miao2018}).

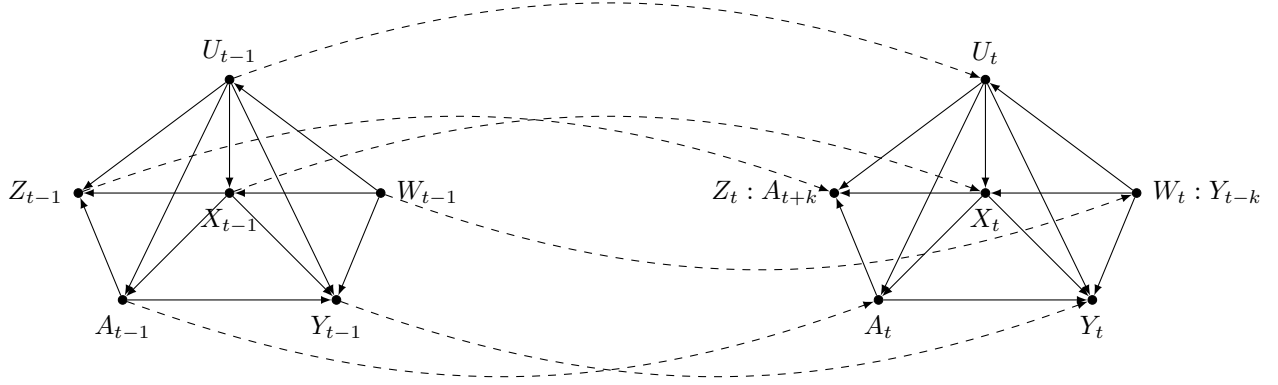
\begin{figure}[H]
\centering
\begin{tikzpicture}[node distance = 2cm]
        \node (x) [label = below:$X_t$, point];
        \node (w) [label = right:$W_t:Y_{t-k}$, point, right of = x, node distance = 2cm];
        \node (y) [label = below:$Y_t$, point, below right of = x, node distance = 2cm];
        \node (u) [label = above:$U_t$, point, above of = x, node distance = 1.5cm];
        \node (a) [label = below:$A_t$, point, below left of = x, node distance = 2cm];
        \node (z) [label = left:$Z_t:A_{t+k}$, point, left of = x, node distance = 2cm];
        \node (x2) [label = below:$X_{t-1}$, point, left of = x, node distance = 10cm];
        \node (w2) [label = right:$W_{t-1}$, point, left of = w, node distance = 10cm];
        \node (y2) [label = below:$Y_{t-1}$, point, left of = y, node distance = 10cm];
        \node (u2) [label = above:$U_{t-1}$, point, left of = u, node distance = 10cm];
        \node (a2) [label = below:$A_{t-1}$, point, left of = a, node distance = 10cm];
        \node (z2) [label = left:$Z_{t-1}$, point, left of = z, node distance = 10cm];
        \path (u) edge  (z);
        \path (u) edge  (x);
        \path (w) edge  (u);
        \path (a) edge  (z);
        \path (x) edge  (z);
        \path (x) edge  (a);
        \path (w) edge  (x);
        \path (x) edge  (y);
        \path (w) edge  (y);
        \path (a) edge  (y);
        \path (u) edge  (a);
        \path (u) edge  (y);

        \path (u2) edge  (z2);
        \path (u2) edge  (x2);
        \path (w2) edge  (u2);
        \path (a2) edge  (z2);
        \path (x2) edge  (z2);
        \path (x2) edge  (a2);
        \path (w2) edge  (x2);
        \path (x2) edge  (y2);
        \path (w2) edge  (y2);
        \path (a2) edge  (y2);
        \path (u2) edge  (a2);
        \path (u2) edge  (y2);
        
        \path[dashed] (u2) edge[bend left  = 20]  (u);
        \path[dashed] (x2) edge[bend left  = 20]  (x);
        \path[dashed] (a2) edge[bend right  = 20]  (a);
        \path[dashed] (y2) edge[bend right  = 20]  (y);
        \path[dashed] (z2) edge[bend left  = 20]  (z);
        \path[dashed] (w2) edge[bend right  = 20]  (w);
\end{tikzpicture}

\caption{A DAG example with time-varying endogenous treatments and time-varying proxies, where $A$ is the treatment, $Y$ is the outcome, $U$ is the unmeasured confounder, and $X,Z,W$ are different type of proxies. }
\label{fig:dag}
\end{figure}

Mobile health (mHealth) studies are one of the most common examples of intensive longitudinal studies (\cite{potter2023}). By enabling real-time monitoring of individual participants and delivering personalized interventions, mHealth technologies have become increasingly valuable for advancing public health. These mHealth studies, often composed of multiple N-of-1 observational studies, face complex time-varying observed and unobserved confounding where confounders both affect treatment decisions and are themselves influenced by past treatments (\cite{neto2022}). An example is illustrated in Figure \ref{fig:dag}. In these observational N-of-1 settings, each individual contributes a rich time series of exposures and outcomes, and there is substantial heterogeneity in trajectories and treatment responses across  (\cite{raman2018}). Researchers are therefore interested in individualized causal effects that vary from person to person, which naturally motivates methodologies that operate directly on within-person time series in mobile health studies rather than aggregating to a single population-level effect.

Various approaches have been proposed to estimate causal effects in time series data. One of them is marginal structural models (MSMs) that leverage the inverse probability of treatment weighting and balance the data such that the effect of time-varying confounding is removed. Its limitations are that MSMs can only balance on known covariates and the number of balancing variables is limited by sample size. Another popular approach is the g-formula. For example, \cite{cai2025} proposed to use the g-formula with state space modeling to estimate dynamic and individualized causal effects in non-stationary time series data; \cite{fowler2025} used an adapted autoregressive hidden Markov model to identify the latent variable as an effect modifier for the causal effect of interest and then obtained g-formula estimators. But their performances rely on correct model specification and the untestable assumption of no unmeasured confounding. Different from MSMs and g-formula estimators, estimators based on deep learning models such as the counterfactual recurrent network (CRN) have been proposed to study dynamic causal effects in time series data. CRN learns representations that balance the distribution of the treatment and control groups, and estimates causal effects over time while handling the bias from time-varying confounders (\cite{bica2020}). Like the methods above, it requires sequential ignorability assumption (no unmeasured confounding and positivity) to hold. There exist methods that could account for unmeasured confounding in time series, such as difference-in-differences (DiD) and instrumental variable. However, in N-of-1 time series data, DiD could be biased when individuals' health states change over time and thus the parallel trends assumption is violated. Standard DiD also could not handle time-varying confounding and autocorrelation. Similarly, instrumental variable methods can be biased when the instrument is weak and sample size is small.

On the other hand, PCI offers an alternative solution that relaxes the assumption of no unmeasured founding. Recent developments in PCI theory (\cite{tchetgen2024}; \cite{ying2024}) have established sufficient conditions for identification in both point treatment and time-varying treatment settings under unmeasured confounding. Semiparametric theory for proximal estimation, along with efficiency theory, have been developed for these settings (\cite{cui2024}). However, despite this progress, existing work for the time-varying treatment setting has primarily focused on joint causal effects over time in longitudinal settings. The contemporaneous or lagged causal effect estimation in time series of N-of-1 studies remains largely unexplored. Our paper extends the PCI framework to address this gap by targeting individual-level contemporaneous and lagged effects in intensive time series. In a motivating multi-year smartphone study of patients with bipolar disorder, prior work using state space models under the assumption of no unmeasured confounding reported a statistically significant contemporaneous effect of text messaging on negative mood for a single participant within a prespecified time window. Yet in this setting, it is plausible that time-varying latent factors such as underlying mood or symptom severity influence both engagement in phone-based communication and subsequent self-reported mood, so that unmeasured confounding may bias individualized effect estimates. Methodologically, this work makes four main contributions. First, it formalizes a proximal causal inference framework for estimating individual contemporaneous and lagged treatment effects in intensive longitudinal N-of-1 time series, extending existing PCI theory beyond joint effects over time. Second, it shows how past outcomes and future treatments, without the need of identifying additional proxy variables, can be exploited as natural proxies for time-varying unmeasured confounders and derives corresponding identification conditions via outcome confounding bridge functions. Third, it develops a class of proximal outcome regression estimators under marginal structural mean models, together with influence-function-based inference and asymptotic variance formulas. Fourth, through simulation studies calibrated to the bipolar smartphone study and an application to the Bipolar Longitudinal Study itself, it demonstrates that accounting for unmeasured confounding via proxies can attenuate previously reported text-messaging effects and materially change substantive conclusions at the individual level. Our approach thus leverages the temporal richness of intensive longitudinal mHealth data to adjust for hidden time-varying confounding and to provide more credible individualized causal effect estimates in real-world observational settings.

The rest of the paper is organized as follows. Notation and the estimand of the contemporaneous and lagged effect used are given in Section 2. In Section 3, we then introduce the proximal causal inference framework and the assumptions needed for identifiability. We also propose a class of proximal outcome regression (POR) estimators and derive influence functions under the class of marginal structural mean models (MSMMs). In Section 4, we evaluate the efficacy and accuracy of our estimator through simulation studies. We consider a scenario that is similar to our data application and cases when some of the identification assumptions are violated. In section 5, we apply our estimator to a multi-year smartphone observational study of bipolar patients and estimate the dynamic eﬀect of phone-based communication on mood of patients with bipolar disorder in an N-of-1 setting. We end the paper with a discussion in Section 6.

\section{Preliminaries}

\subsection{Notation}

We focus on a single participant observed over $T$ time points, as the target estimand in this paper is an individualized contemporaneous or lagged causal effect in an N-of-1 setting. Let $Y_{t}, A_{t}, \mathbf{L}_{t}$, be the outcome, treatment, and covariates at time $t\in \{1,2,...,T\}$, respectively, for one single subject. We assume $A_t$ is binary, and $Y_{t}, \mathbf{L}_{t}$ are continuous. We define $\{\mathbf{L}_{1:T}=(\mathbf{L}_{1}, ..., \mathbf{L}_{T}), A_{1:T}=(A_1,...,A_T), \text{ and }Y_{1:T}=(Y_1,...,Y_T)\}$. We denote $Y_t(A_{t}=a_{t})$ the potential outcome at time $t$ that would have been observed under the treatment $A_{t}=a_t$. Our goal is to estimate either the contemporaneous effect of $A_t$ on $Y_t$, or the $q-$lagged effect of $A_{t-q}$ on $Y_t$ in the presence of time-varying unmeasured confounding. We use $U_{1:T}=(U_1,...,U_T)$ to denote time-varying unmeasured variables that confound the treatment-outcome relationship over the outcome.

Because of time-varying unmeasured confounding, the sequential randomization assumption (SRA) that is essential to the identification of causal effects in longitudinal studies often fails. However, the PCI framework can help relax SRA by explicitly incorporating measured covariates as proxies of underlying confounding mechanisms longitudinally. To formally introduce the longitudinal PCI framework, suppose that the observed covariates $\mathbf{L}_{1:T}$ consist of three types $(X_{1:T}, Z_{1:T}, W_{1:T})$ where $X_{1:T}$ are common causes of subsequent treatment and outcome variables; $Z_{1:T}$ are referred to as sequential treatment-inducing proxies; $W_{1:T}$ are referred to as sequential outcome-inducing proxies (\cite{tchetgen2024}). A treatment-inducing proxy is related with the outcome only through confounders. A outcome-inducing proxy is related with the treatment only through confounders. For a list of directed acyclic graphs for proxy, treatment, and outcome relationships, please review the supplementary material of \cite{tchetgen2024}. 

In this paper, we focus on estimating the contemporaneous and $q$-lagged causal effects. In terms of the contemporaneous effect, we are estimating the given effect of an intervention $A$ at time $t$ on the contemporaneous outcome $Y$ measured within the same unit of time. We use the similar definitions in \cite{fowler2025}. Specifically, \[\mathbb{E}\{Y_t(a_t=1)-Y_t(a_t=0)|V\},\] where $V \subset X_{t-j:t}\cup A_{t-j:t-1}$.

Similarly, for $q$-lagged effect, we want to evaluate the effect of a past intervention at time $t - q$ on
a future outcome at time $t$ for some $q \geq 1$:\[\mathbb{E}\{Y_t(a_{t-q}=1)-Y_t(a_{t-q}=0)|V\},\] where $V \subset X_{t-j:t-q}\cup A_{t-j:t-q-1}$.

\subsection{Estimation in the absence of unmeasured confounding}


For inference, we follow the approach in \cite{ying2024} and consider the class of marginal structural mean models (MSMMs). The marginal structural models (MSMs) can encode the causal effects of time-varying treatment subject to time-varying confounding (\cite{robins2000}). Different from MSMs, MSMMs place an additional restriction on the mean of the counterfactual variables possibly conditional on the baseline variables. In the case of contemporaneous causal effect, we restrict the mean of $Y_t(a_t)$ conditional on the past history variables $V \subset X_{t-j:t}\cup A_{t-j:t-1}$ such that 
\[\mathbb{E}\{Y_t(a_t)|V\}=g(a_{t}, V;\beta) \text{ for any } a_t\in \mathcal{A},\] for a known function $g()$ and $p$-dimensional parameter $\beta$. For example, a simple working choice is $g(a_{t}, V;\beta)=\beta_0+\beta_1a_t+\beta_2V$, although richer specifications with a different choice of link function are allowed.

We focus on the contemporaneous effect in the main text. But we can replace $Y_t(a_t)$ with $Y_t(a_{t-q})$ when studying $q$-lagged effect, and follow the same procedure conditioning on the past history variables happening before $t-q$. We are interested in estimating $\beta$ and denote the truth as $\beta_*$. This formulation yields valid causal inference under the assumptions of consistency, positivity, sequential ignorability, stationarity, and correct model specification. Here, sequential ignorability assumption includes the no unmeasured confounding condition, which we aim to relax in our setting. When sequential ignorability holds, $\mathbb{E}\{Y_t(a_t)\}$ can also be nonparametrically identified by the longitudinal g-formula (\cite{robins1997}).

\section{PCI framework}

In this section, we introduce new assumptions for proximal identification of the counterfactual mean $\mathbb{E}\{Y_t(a_t)|V\}$. The identification is based on outcome confounding bridge functions defined as a solution to certain Fredholm integral equations
of the first kind (\cite{tchetgen2024}).Then we describe a class of estimating equations for the parameters of an MSMM, and provide the details about estimation and inference results.


\subsection{Assumptions}

We start with a longitudinal generalization of the standard assumptions about consistency and positivity. Assumption 1 links counterfactual outcomes to observed outcomes. Assumption 2 essentially states that for any realized history of treatment and covariates from time $t-j$ ($j\in \mathbb{Z}\geq 1$) to time $t-1$, there is a non-negligible chance that subjects will receive either treatment. Because the methods in Section 2.2 relies on sequential ignorability, including no unmeasured confounding, that assumption may fail in the presence of time-varying unmeasured confounding. However, the PCI framework can relax SRA by explicitly incorporating measured covariates as proxies of underlying confounding mechanisms longitudinally.

\noindent \textbf{Assumption 1} Consistency:
\[Y_t=Y_t(A_t) \text{ almost surely}.\] 

\noindent \textbf{Assumption 2} Positivity:
\[0<Pr(A_t=a|A_{t-j:t-1}, U_{t-j:t}, L_{t-j:t})<1 \text{ for } a=1,0 \text{ almost surely}.\] 

Then we would like to introduce the rest of our identifying assumptions for the contemporaneous causal effect in the time series setting. These assumptions rely on the conditions introduced in \cite{tchetgen2024} and \cite{ying2024}. 

\noindent \textbf{Assumption 3} Proxy independence:

\[Y_t\indep Z_t |A_t=a_t,A_{t-j:t-1},X_{t-j:t},U_{t-j:t},\] \[W_t\indep (Z_t,A_t) |X_{t-j:t},U_{t-j:t}.\] This assumption defines the sequential proxies and implies certain independence relationships. Specifically, it states that treatment-inducing proxy $Z_t$ does not directly cause $Y_t$ after intervening on $A_t$, and outcome-inducing proxy $W_t$ is associated with $Z_t$ and $A_t$ only through $X_{t-j:t},U_{t-j:t}$.

In the literature of proximal causal inference, there exist two identification approaches based on different bridge functions. We mainly focus on the identification result via outcome confounding bridge functions. They are defined as a solution to certain Fredholm integral equations of the first kind (\cite{tchetgen2024}, \cite{miao2018}).

\noindent \textbf{Assumption 4} Bridge function:

There exists a function $h(w_{t-j:t},a_{t-j:t},x_{t-j:t})$ such that, \[E[Y_t|Z_{t-j:t},A_{t-j:t}, X_{t-j:t}]=\int h(w_{t-j:t},a_{t-j:t},x_{t-j:t})dF(w_{t-j:t}|Z_{t-j:t},A_{t-j:t},X_{t-j:t}).\]

Although this assumption does not have a unique solution for the bridge function, any bridge function satisfying the condition can uniquely identify the true contemporaneous causal effect (\cite{tchetgen2024}).  

\noindent \textbf{Assumption 5} Completeness:

(a)For any $a_{t-j:t}, x_{t-j:t}$, and any square-integrable function $v$,
$$E[v\{U_t\}|a_{t-j:t}, Z_{t-j:t}, x_{t-j:t}]=0 \text{ if and only if } v\{U_t\}=0 \text{ almost surely.}$$

(b)For any $a_{t-j:t}, x_{t-j:t}$, and any square-integrable function $v$,
$$E[v\{Z_t\}|a_{t-j:t}, W_{t-j:t}, x_{t-j:t}]=0 \text{ if and only if } v\{Z_t\}=0 \text{ almost surely.}$$

This completeness assumption requires the proxies have sufficient variability relative to that of unmeasured confounding. Intuitively, in categorical data, both proxies must have at least as many categories as the unmeasured confounders. As noted in \cite{tchetgen2024}, the completeness assumption is widely satisfied in common distributions, even though it is untestable. We also present the results from simulation studies assessing the sensitivity of proximal causal inference to violation of this assumption in our time series setting. 

\subsection{Identification formula}

\textbf{Theorem 1}

Under assumptions (1-5), we have that \[\mathbb{E}(Y_t(a_t)|u_{t-j:t},a_{t-j:t},x_{t-j:t})=\mathbb{E}(h(w_{t-j:t},a_{t-j:t},x_{t-j:t})|u_{t-j:t},a_{t-j:t},x_{t-j:t}).\]

It follows that  \[\mathbb{E}(Y_t(a_t)|a_{t-j:t},x_{t-j:t})=\mathbb{E}(h(w_{t-j:t},a_{t-j:t},x_{t-j:t})|a_{t-j:t},x_{t-j:t}),\]

and therefore \[\mathbb{E}\{Y_t(a_t)|V\}=\mathbb{E}\{h(w_{t-j:t},a_{t-j:t},x_{t-j:t})|V\}.\]

\noindent\textit{proof}

From assumption 4 and by law of total expectation and assumption 3, we have 
\begin{align*}
    &\mathbb{E}[\mathbb{E}\{Y_t|u_{t-j:t},a_{t-j:t},x_{t-j:t}\}|z_{t-j:t},a_{t-j:t},x_{t-j:t}]\\
    &=\mathbb{E}(Y_t|z_{t-j:t},a_{t-j:t},x_{t-j:t})\\
    &=\mathbb{E}\{h(w_{t-j:t},a_{t-j:t},x_{t-j:t})|z_{t-j:t},a_{t-j:t},x_{t-j:t}\}\\
    &\mathbb{E}\{\mathbb{E(}h(w_{t-j:t},a_{t-j:t},x_{t-j:t})|u_{t-j:t},a_{t-j:t},x_{t-j:t})|z_{t-j:t},a_{t-j:t},x_{t-j:t}\}.
\end{align*}

By Assumption 5, we have \[\mathbb{E}(Y_t(a_t)|u_{t-j:t},a_{t-j:t},x_{t-j:t})=\mathbb{E}(h(w_{t-j:t},a_{t-j:t},x_{t-j:t})|u_{t-j:t},a_{t-j:t},x_{t-j:t}).\]

By simply taking an expectation over the unmeasured confounders, we have \[\mathbb{E}(Y_t(a_t)|a_{t-j:t},x_{t-j:t})=\mathbb{E}(h(w_{t-j:t},a_{t-j:t},x_{t-j:t})|a_{t-j:t},x_{t-j:t}),\] and \[\mathbb{E}\{Y_t(a_t)|V\}=\mathbb{E}\{h(w_{t-j:t},a_{t-j:t},x_{t-j:t})|V\}.\]

\subsection{Proximal estimation under MSMM}

We have shown that the contemporaneous causal effect can be identified nonparametrically in the presence of unmeasured confounding. In this part, we follow the approach in \cite{ying2024} and derive the set of estimators under a semiparametric MSMM. The MSMM's restriction in section 2.2 gives us that for any $p$-dimensional measurable function $d$,
\begin{equation} \label{estequ}
\mathbb{E}\left\{\sum_{a_{t-j:t} \in \mathcal{A}} d(a_{t-j:t}, V)\left[\mathbb{E}\{Y_t(a_t)|V\}-g(a_{t-j:t}, V ; \beta)\right]\right\}=0.
\end{equation}
Under Theorem 1, we can have the following equivalent moment equation, and our estimator for $\beta$ is defined as the solution to it \[\mathbb{E}[D(\beta, d, b)]=0,\] where $D(\beta, d, b)=\sum_{a_{t-j:t} \in \mathcal{A}} d(a_{t-j:t}, V)\left[h\left(a_{t-j:t} ; b\right)-g(a_{t-j:t}, V ; \beta)\right]$; $b$ is the parameter of the bridge function.

However, deriving the estimator from $\mathbb{P}_n[D(\beta, d, b)]=0$ is challenging because it depends crucially on complicated functions of the true observed data distribution that is unknown. Empirical solutions to these integral equations are notoriously challenging to compute due to the ill-posedness nature of the problem, typically requiring a form of regularization. Thus, we consider a practical approach to constructing feasible moment estimating equations for $\beta_*$ under low-dimensional smooth working models for the nuisance parameter $h$. Let $h(\cdot)=h(\cdot; b)$  parametric working models indexed by low-dimensional parameter $b$. Let $M=m(A_{t-j:t}, Z_{t-j:t}, X_{t-j:t})$ for some measurable function $m()$ that is of the same dimension as $b$. By Theorem 2, the estimator of $h$ can be obtained by solving the following estimating equation \[\mathbb{P}_n\{[Y-h(a_{t-j:t})]M\}=0.\]

In our simulation studies, we use simple working models as the following:
\[h(a_{t-j:t;b})=h(1, A_{t-j:t}, W_{t-j:t}, X_{t-j:t};b)=b_{0}+b^T_{a}A_{t-j:t}+b^T_{w}W_{t-j:t}+b^T_{x}X_{t-j:t}=\mathbf{R}b,\]

\[M=m(A_{t-j:t}, Z_{t-j:t}, X_{t-j:t})=(1, A_{t-j:t}, Z_{t-j:t}, X_{t-j:t})^T.\]

\noindent \textbf{Theorem 2} (\cite{ying2024})

Let $\mathcal{M}_h$ denote the collection of data generating laws under working models $h(\cdot;b)$.
Influence functions of regular and asymptotically estimators of $b$ under $M_h$ are of the form: \[\left(\mathbb{E}\left[\frac{\partial h(a_{t-j:t};b)}{\partial b}M\right]\right)^{-1}((Y-h(a_{t-j:t}))M)^T.\]

\textit{proof}

Under the assumed models, we have shown from Theorem 1's proof that 
    \[\mathbb{E}_s(Y_t|z_{t-j:t},a_{t-j:t},x_{t-j:t})=\mathbb{E}_s\{h(a_{t-j:t};b_s)|z_{t-j:t},a_{t-j:t},x_{t-j:t}\},\]
under any regular parametric submodel $\{p_s:s\}$.
Furthermore, this moment condition is equivalent to \[\mathbb{E}_s[(Y_t-h(w_{t-j:t},a_{t-j:t},x_{t-j:t}))M]=0.\]
Differentiating at $s=0$ yields \[\mathbb{E}\left[\frac{\partial h(a_{t-j:t};b_s)}{\partial b}M\right]\left.\frac{\partial b_s}{\partial s}\right|_{s=0}=\mathbb{E}[(Y_t-h(w_{t-j:t},a_{t-j:t},x_{t-j:t}))MS(\mathcal{O})].\]
Therefore, we have \[\left.\frac{\partial b_s}{\partial s}\right|_{s=0}=\mathbb{E}(\left(\mathbb{E}\left[\frac{\partial h(a_{t-j:t};b)}{\partial b}M\right]\right)^{-1}((Y-h(a_{t-j:t}))M)^TS(\mathcal{O})).\]
\subsection{Asymptotic distribution and influence function}

We now use the estimator of the bridge function $\hat{h}$ to construct a corresponding substitution estimators of $\beta_*$. Suppose that $n^{1 / 2}\left(\hat{b}-b^{*}\right)=n^{-1 / 2} \sum_{i=1}^{n} \varepsilon_{b^{*}, i}+o_{p}(1)$, where $b^{*}$ is the true value of the bridge-function parameter and this expansion holds under $\mathcal{M}_{h}$, the collection of observed data generating laws under which the working model $h(\cdot;b)$ is correctly specified. The class of proximal outcome regression (POR) estimators $\hat{\beta}$ that can be obtained by solving

$$
n^{1 / 2}\left\{D\left(\beta, d, \hat{b}\right)\right\}=n^{1 / 2} \mathbb{P}_{n}\left\{\sum_{a_{t-j:t} \in \mathcal{A}} d(a_{t-j:t}, V)\left[\widehat{h}\left(a_{t-j:t} ; \hat{b}\right)-g(a_{t-j:t}, V ; \beta)\right]\right\}=o_{p}(1)
$$

is consistent and asymptotically normal for $\beta_{*}$, with influence function $(\operatorname{IF})$ given by

\begin{equation} \label{inffun}
\begin{aligned}
 \sqrt{n}(\hat{\beta}-\beta_*)&=\frac{1}{\sqrt{n}}\sum^n_{i=1}\text{IF}_i+o_p(1) \\
& =\frac{1}{\sqrt{n}}\sum^n_{i=1}\left[-\mathbb{P}(\frac{\partial D(\beta_*,d,b^*)}{\partial\beta})\right]^{-1}\left[D_i(\beta_*,d,b^*)+\mathbb{P}(\frac{\partial D(\beta_*,d,b^*)}{\partial b})\varepsilon_{b^*,i}\right]+o_p(1).
\end{aligned}
\end{equation}

\textit{proof}

Consistency and asymptotic normality of $\hat{\beta}$ follows immediately from standard law of large numbers and central limit theorems. Please refer to Theorem B.4. in the supplementary materials of \cite{ying2024}. Therefore, we need only to show the influence function form of $\hat{\beta}$.

Under some regularity conditions, (3.10) in \cite{newey1994} shows that influence function of $\hat{\beta}$ can be derived from $\mathbb{P}[D(\beta_*, d, b^*)]=0$, and it takes the following form 
\[\left[-\partial\mathbb{P}(\frac{ D(\beta_*,d,b^*)}{\partial\beta})\right]^{-1}\left[D(\beta_*, b^*)-\frac{\partial}{\partial b}\mathbb{P}_n[D(\beta_*, b^*)]\text{IF}_{b}\right].\]

Because $\text{IF}_{b}=\varepsilon_{b^{*}}$, the influence function is \[\left[-\mathbb{P}(\frac{\partial D(\beta_*,d,b^*)}{\partial\beta})\right]^{-1}\left[\sum^n_{i=1}(D_i(\beta_*, b^*)-\mathbb{P}(\frac{\partial D(\beta_*, b^*)}{\partial b})\varepsilon_{b^{*},i})\right].\]

Using this expression, we obtain an explicit form of the individual influence function, which in turn yields the asymptotic distribution of our estimator.

\section{Simulation Studies}

In this section, to assess the performance of our proposed method, we present the results from simulation studies. We first consider a setting where data is generated to match the setting of our application. Then we investigate our method's robustness when some of the assumptions are violated.

\subsection{Data Generating Process}

\begin{figure}[H]
\centering
\begin{subfigure}[b]{1\linewidth}
\begin{tikzpicture}[node distance = 2cm]
        \node (x) [label = below:$X_t$, point];
        \node (w) [label = right:$W_t:Y_{t-k}$, point, right of = x, node distance = 2cm];
        \node (y) [label = below:$Y_t$, point, below right of = x, node distance = 2cm];
        \node (u) [label = above:$U_t$, point, above of = x, node distance = 1.5cm];
        \node (a) [label = below:$A_t$, point, below left of = x, node distance = 2cm];
        \node (z) [label = left:$Z_t:A_{t+k}$, point, left of = x, node distance = 2cm];
        \node (x2) [label = below:$X_{t-1}$, point, left of = x, node distance = 10cm];
        \node (w2) [label = right:$W_{t-1}$, point, left of = w, node distance = 10cm];
        \node (y2) [label = below:$Y_{t-1}$, point, left of = y, node distance = 10cm];
        \node (u2) [label = above:$U_{t-1}$, point, left of = u, node distance = 10cm];
        \node (a2) [label = below:$A_{t-1}$, point, left of = a, node distance = 10cm];
        \node (z2) [label = left:$Z_{t-1}$, point, left of = z, node distance = 10cm];
        \path (u) edge  (z);
        \path (u) edge  (x);
        \path (w) edge  (u);
        \path (a) edge  (z);
        \path (x) edge  (z);
        \path (x) edge  (a);
        \path (w) edge  (x);
        \path (x) edge  (y);
        \path (w) edge  (y);
        \path (a) edge  (y);
        \path (u) edge  (a);
        \path (u) edge  (y);

        \path (u2) edge  (z2);
        \path (u2) edge  (x2);
        \path (w2) edge  (u2);
        \path (a2) edge  (z2);
        \path (x2) edge  (z2);
        \path (x2) edge  (a2);
        \path (w2) edge  (x2);
        \path (x2) edge  (y2);
        \path (w2) edge  (y2);
        \path (a2) edge  (y2);
        \path (u2) edge  (a2);
        \path (u2) edge  (y2);
        
        \path[dashed] (u2) edge[bend left  = 20]  (u);
        \path[dashed] (x2) edge[bend left  = 20]  (x);
        \path[dashed] (x2) edge[bend left  = 20]  (u);
        \path[dashed] (a2) edge[bend left  = 20]  (u);
        \path[dashed] (a2) edge[bend right  = 20]  (x);
        \path[dashed] (a2) edge[bend right  = 20]  (a);
        \path[dashed] (u2) edge[bend right  = 30]  (a);
        \path[dashed] (x2) edge[bend right  = 20]  (a);
        \path[dashed] (a2) edge[bend right  = 20]  (y);
        \path[dashed] (y2) edge[bend left  = 10]  (u);
        \path[dashed] (y2) edge[bend left  = 20]  (x);
        \path[dashed] (y2) edge[bend right  = 20]  (y);
        \path[dashed] (z2) edge[bend left  = 20]  (z);
        \path[dashed] (w2) edge[bend right  = 20]  (w);
\end{tikzpicture}
\end{subfigure}%

\caption{Data Generating Process}
\label{fig:dgp}
\end{figure}
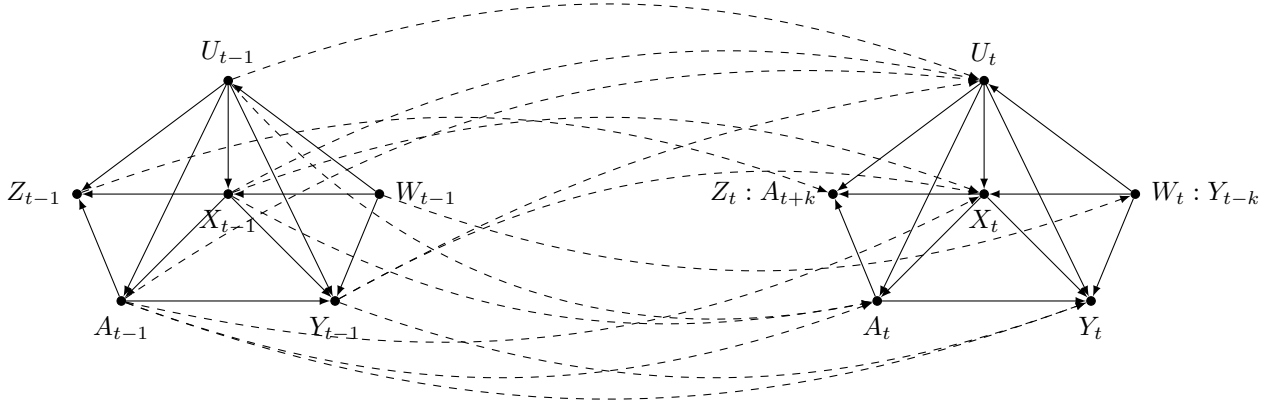

We simulate datasets with T= 90 time points based on Figure \ref{fig:dgp}. We generate \(U_t\) from a Bernoulli model with a logit link depending on \(U_{t-1}\), \(Y_{t-1}\), \(X_{t-1}\), and \(A_{t-1}\). We then generate \(X_t\) as a linear function of \(X_{t-1}\), \(Y_{t-1}\), \(A_{t-1}\), and \(U_t\), and generate \(A_t\) from a Bernoulli model with a logit link depending on \(A_{t-1}\), \(U_t\), \(X_t\), \(U_{t-1}\), and \(X_{t-1}\). Finally, \(Y_t\) is generated as a linear function of \(Y_{t-1}\), \(A_{t-1}\), \(A_t\), \(U_t\), and \(X_t\), with Gaussian noise. 
We focus on estimation of the contemporaneous effect $A_t$ on $Y_t$ that we set as $\beta=-2.0$, which is numerically close to the observed effect size in the data application. Additional simulation scenarios with longer follow-up lengths are reported in the supplementary materials.

We also evaluate the robustness of PCI to the violations of the completeness, proxy independence, proxy relevance assumptions. To simulate the violation of the completeness assumption, we change the unmeasured confounder $U$ to be continuous by replacing the Bernoulli model with a linear function with Gaussian noises. The completeness assumption requires both proxies to have sufficient variability relative to variability of $U$, but binary proxies have less variability than continuous unmeasured confounder. For the violation of the proxy independence assumption, we simply allow $Y$ to affect $A$ such that the outcome-inducing proxy $W$ is not conditionally independent of $A$. The last violation case, we decrease the autocorrelation in the data. We make proxies $Z, W$ to be not $U$-relevant in the sense that they are independent of $U$ conditional on $A, X$ and the other proxy. The $U$-relevance is part of the completeness assumption, but we want to discuss it separately because it has been emphasized in many literature on proximal causal inference and a strong autocorrelation is a key component to apply PCI in the time series setting.

\subsection{Numerical Results}
We run the PCI estimation procedure 1000 times under each scenario considered, and report the average estimates, 95\% confidence intervals (average upper and lower bounds), and coverages. We present the results of $T=90$ in this section and those of $T=10000$ can be found in the appendix. For each scenario, we also show how the estimates vary based on the values of $k$ and $j$, where $j$ determines how much of history data we condition on, and $k$ is used to define our proxies. For example, when $k=2$ and $j=3$, we set $Z_t(A_{t+2})$ and $W_t(Y_{t-2})$ as the proxies corresponding to $A_t$ and $Y_t$, and we additionally adjust for history data from $t-3$ up to $t$ including $Z_{t-3:t-1}, W_{t-3:t-1}, A_{t-3:t-1}, $ and $X_{t-3:t}$. We omit the results when $j\geq k$, because in these cases the matrix inside the inverse term of the influence function \ref{inffun} would be singular and thus we cannot construct the confidence intervals. 

Table \ref{tab1} summarized the simulation results for four different scenarios. The PCI estimator is approximately unbiased across all choices of $k$ and $j$, with estimates ranging from -1.950 to -1.876 around the true value of -2.0, when all assumptions are satisfied. The 95\% confidence intervals are wide but uniformly achieve nominal or near-nominal coverage, with coverage ranging from 99.1\% to 99.9\%. For fixed $j$, using proxies closer to the current time point, corresponding to smaller $k$, generally yields estimates closer to the truth. Likewise, conditioning on a richer recent history, corresponding to larger $j$, tends to slightly improve estimation and maintain excellent coverage. 



When the completeness assumption is violated, the PCI estimator remains close to the truth in some settings but becomes slightly more variable and tends to overestimate the effect, especially for larger $k$. Coverage is still great when $k=2$, but it drops to 72.7\%–82.8\% for $k=3$ and $k=4$, indicating that the intervals are too narrow relative to the sampling variability. The strongest performance occurs at $k=2,j=1$, where the proxies are closest in time to the target treatment and outcome and appear to be more strongly correlated with the latent confounding structure. 

When the proxy independence assumption is violated, the PCI estimator exhibits noticeably more bias and substantially wider confidence intervals than in the correctly specified setting. The effect estimates are most distorted when $j=1$ and $k=2$, where the estimated effect is -2.547 with a very wide confidence interval (-13.901,8.807), although coverage remains high because of the inflated interval width. As more recent history is included, the bias is attenuated and coverage improves, with the best performance occurring at $j=3,k=4$ where the estimate is closest to the truth and coverage returns to 99.6\%. The linear regression estimators also perform poorly here, with much lower coverage and persistent bias even after adding lagged variables, suggesting that simple lag adjustment does not fully address this type of violation.

When the proxy relevance assumption is violated, the PCI estimator is still approximately unbiased at $T=90$, but the confidence intervals become very wide, reflecting weak information from the proxies. Coverage remains close to nominal in all settings, but the large interval lengths indicate a loss of efficiency rather than bias. This pattern is consistent with the proxies being weakly related to the unmeasured confounder, so the estimator remains stable in the short series but becomes much less informative when the follow-up is longer ($T=10000$), as shown in the supplementary materials. In that longer setting, coverage deteriorates substantially for the linear models without the strongest adjustment, reinforcing that weak proxy relevance mainly hurts precision and finite-sample stability.


In addition, we compare PCI with three linear model-based alternatives. The first regresses the outcome on the current exposure and observed covariates only($Y_t= \beta_0+\beta_1A_t+\beta_2X_t$). The second adds first-order lagged covariates, lagged outcomes, and lagged treatments($Y_t= \beta_0+\beta_1A_t+\beta_2X_t+\beta_3W_t+\beta_4X_{t-1}+\beta_5W_{t-1}+\beta_6A_{t-1}$). The third further includes second-order lagged covariates, outcomes, and treatments($Y_t= \beta_0+\beta_1A_t+\beta_2X_t+\beta_3W_t+\beta_4X_{t-1}+\beta_5W_{t-1}+\beta_6A_{t-1}+\beta_{7}X_{t-2}+\beta_{8}W_{t-2}+\beta_{9}A_{t-2}$). The results are reported in Table \ref{tab2}. These models provide a useful benchmark for assessing how much adjustment can be gained from including lagged history when proximal proxies are unavailable or difficult to justify. Table \ref{tab2} shows that adding lagged history improves the performance of the linear models relative to the simplest specification, but these linear model estimators still exhibit non-negligible bias and lower coverage than PCI in several settings. In contrast, PCI remains closer to the truth and achieves better calibrated uncertainty quantification, especially when valid proximal proxies are available.


Taken together, these simulation scenarios offer guidance for applying the PCI framework in intensive time series. When investigators have strong reasons to suspect time-varying unmeasured confounding and can identify plausible proxies (e.g. future treatments and past outcomes with clear temporal structure), PCI is a natural choice because it yields approximately unbiased estimates and nominal coverage when its identification assumptions hold or are only mildly violated. In contrast, when suitable proxies are difficult to justify or do not have sufficient variability relative to the variability of the unmeasured confounder, our results suggest that PCI can become unstable and may overestimate effects, in which case linear models with carefully chosen lags may be preferable. Across settings, increasing $j$ to condition on a richer recent history generally improves bias and coverage by better capturing the dependence structure, but very large $j$ may inflate variance and create numerical instability, so we recommend starting from small $j$ (e.g., 1–2 lags) and increasing only when there is substantive or empirical evidence of longer memory in the process.

\begin{table}[H] 
\resizebox{\linewidth}{!}{%
\begin{tabular}{ccccccc}
        \toprule
        & \multicolumn{2}{c}{j=1} & \multicolumn{2}{c}{j=2} & \multicolumn{2}{c}{j=3} \\
        \cmidrule(r){2-3} \cmidrule(r){4-5} \cmidrule(r){6-7}
        k & $\hat{\beta}$ (95\% CI) & Coverage & $\hat{\beta}$ (95\% CI) & Coverage & $\hat{\beta}$ (95\% CI) & Coverage \\
        \midrule
          & \multicolumn{6}{l}{\textbf{No violation (baseline)}} \\
        \midrule
        2 & -1.950 (-21.402, 17.502) & 99.6 & - &  & - & \\
        3 & -1.909 (-8.006, 4.189) & 99.1 & -1.922 (-10.344, 6.500) & 99.9 & - & \\
        4 & -1.876 (-7.111, 3.589) & 99.2 & -1.905 (-7.113, 3.303) & 99.9 & -1.917 (-12.907, 9.073) & 99.9 \\
        \midrule
         & \multicolumn{6}{l}{\textbf{Violation to completeness assumption}} \\
        \midrule
        2 & -2.057 (-5.907, 1.793) & 99.4 & - &  & - & \\
        3 & -2.138 (-3.335, -0.942) & 79.8 & -2.088 (-2.444, -1.732) & 72.7 & - & \\
        4 & -2.231 (-3.530, -0.934) & 76.6 & -2.127 (-3.025, -1.230) & 82.8 & -2.105 (-2.595, -1.615) & 77.2 \\
        \midrule
         & \multicolumn{6}{l}{\textbf{Violation to proxy independence assumption}} \\
         \midrule
        2 & -2.547 (-13.901, 8.807) & 99.5 & - &  & - & \\
        3 & -1.285 (-4.717, 2.148) & 73.0 & -1.845 (-14.006, 10.315) & 99.6 & - & \\
        4 & -1.368 (-4.129, 1.393) & 63.7 & -1.486 (-6.346, 3.375) & 85.0 & -1.712 (-13.514, 10.091) & 99.6  \\
        \midrule
         & \multicolumn{6}{l}{\textbf{Violation to proxy relevance assumption}} \\
         \midrule
        2 & -1.912 (-4.756, 0.932) & 99.4 & -  &  & - & \\
        3 & -1.883 (-36.227, 32.461) & 99.3 & -1.902 (-7.205, 3.401) & 99.8 & - & \\
        4 & -1.868 (-5.900, -2.165) & 98.7 & -1.887 (-12.665, 8.892) & 99.5 & -1.900 (-7.003, 3.203) & 99.8 \\
        \bottomrule
    \end{tabular}
}

  \centering 
  \caption{Simulation results for the contemporaneous effect under the main setting with follow-up length $T=90$ and true effect $\beta=-2.0$. We report the average estimated effect $\hat{\beta}$, the average 95\% confidence interval, and empirical coverage for the \textbf{proximal causal inference (PCI) estimator} under the scenario where all assumptions are satisfied and under scenarios violating the completeness, proxy independence, or proxy relevance assumptions. The parameters $k,j$ determine the temporal placement of the proxy variables and the length of recent history included in the model, respectively.} 
  \label{tab1} 
\end{table}

\begin{table}[H] 
\resizebox{\linewidth}{!}{%
\begin{tabular}{ccccccc}
        \toprule
         & \multicolumn{2}{c}{lm} & \multicolumn{2}{c}{lm with lag-1 data} & \multicolumn{2}{c}{lm with lag-2 data} \\
        \cmidrule(r){2-3} \cmidrule(r){4-5} \cmidrule(r){6-7}
         k & $\hat{\beta}$ (95\% CI) & Coverage & $\hat{\beta}$ (95\% CI) & Coverage & $\hat{\beta}$ (95\% CI) & Coverage \\
        \midrule
        & \multicolumn{6}{l}{\textbf{No violation (baseline)}} \\
        \midrule
        2 & -2.116 (-2.703, -1.528) & 92.2 & -1.947 (-2.252, -1.643) & 92.1 & -1.950 (-2.258, -1.643) & 92.8 \\
        \midrule
        & \multicolumn{6}{l}{\textbf{Violation to completeness assumption}} \\
        \midrule
        2 & -0.818 (-1.615, -0.022) & 14.7 & -1.811 (-2.200, -1.421) & 82.7 & -1.855 (-2.236, -1.473) & 87.0 \\
        \midrule
        & \multicolumn{6}{l}{\textbf{Violation to proxy independence assumption}} \\
        \midrule
        2 & -1.028 (-1.382, -0.674) & 0.0 & -1.620 (-1.939, -1.301) & 34.9 & -1.638 (-1.964, -1.312) & 39.4 \\
                \midrule
        & \multicolumn{6}{l}{\textbf{Violation to proxy relevance assumption}} \\
        \midrule
        2 & -2.098 (-2.674, -1.522) & 91.5 & -1.915 (-2.232, -1.597) & 90.8 & -1.918 (-2.239, -1.597) & 91.7 \\
        \bottomrule
    \end{tabular}
}
  \centering 
  \caption{Simulation results for the contemporaneous effect under the main setting with follow-up length $T=90$ and true effect $\beta=-2.0$, based on \textbf{linear model comparison methods}. We report the average estimated effect $\hat{\beta}$, the average 95\% confidence interval, and empirical coverage for models with no lagged history, with one lag, with two lags. Results are shown under the scenarios where the completeness, proxy independence, or proxy relevance assumptions are satisfied or violated.} 
  \label{tab2} 
\end{table}

For lagged effects, Table \ref{tab_lag} shows a similar pattern to the contemporaneous-effect analysis. When all identification assumptions are satisfied, PCI yields estimates that are close to the true lagged effect $\beta=-1.154$ across choices of $k,j$. The estimates vary only modestly, from -1.173 to -1.086, suggesting that the procedure is stable across different choices of proxy timing and conditioning history. The 95\% confidence intervals remain wide, especially when $j$ is small, but empirical coverage stays close to nominal in all settings, ranging from 98.8\% to 99.8\%. Overall, these results indicate that PCI can recover lagged causal effects with good calibration when the proximal identification assumptions hold. More extensive lagged-effect simulations with longer follow-up are provided in the supplementary materials.

\begin{table}[H] 
\resizebox{\linewidth}{!}{%
\begin{tabular}{ccccccc}
        \toprule
        \multicolumn{7}{l}{\textbf{No violation}} \\
        \midrule
         & \multicolumn{2}{c}{j=1} & \multicolumn{2}{c}{j=2} & \multicolumn{2}{c}{j=3} \\
        \cmidrule(r){2-3} \cmidrule(r){4-5} \cmidrule(r){6-7}
         k & $\hat{\beta}$ (95\% CI) & Coverage & $\hat{\beta}$ (95\% CI) & Coverage & $\hat{\beta}$ (95\% CI) & Coverage \\
        \midrule
        3 & -1.170 (-8.860, 6.520) & 99.0 & - &  & - & \\
        4 & -1.174 (-4.112, 1.764) & 99.6 & -1.112 (-5.314, 3.089) & 99.8 & - &  \\
        5 & -1.173 (-4.950, 2.604) & 98.8 & -1.115 (-8.246, 6.017) & 99.7 & -1.086 (-68.702, 66.530) & 99.8  \\
        \bottomrule
    \end{tabular}
}
  \centering 
  \caption{Simulation results for  the lagged effect under the main setting with follow-up length $T=90$ and true effect $\beta=-1.154$. We provide the estimated $\beta$, average 95\% confidence intervals, and coverages from linear models with lagged data.} 
  \label{tab_lag} 
\end{table}

\section{Data Application}

The Bipolar Longitudinal Study is a mHealth cohort study that has enrolled 74 individuals with schizophrenia or bipolar disorder from the Psychotic Disorders Division at McLean Hospital since February 2016. Upon enrollment, each participant completed a comprehensive DSM-5 diagnostic evaluation. Passive sensing via smartphones and wearable fitness trackers captured physical activity, GPS location, and call/text logs, and a customized 5-minute survey was administered daily at 5:00 pm through the Beiwe platform (\cite{huang2019}, \cite{onnela2021}) to assess mood, sleep, social activities, and psychotic symptoms. We evaluate the contemporaneous causal effect of phone-based social connectivity on patients’ self-reported negative mood, building on prior BLS findings (\cite{valeri2023}, \cite{fowler2024}, \cite{cai2022}, \cite{cai2025}). The primary outcome, self-reported negative mood \(Y_t\), is a composite score from 0 (best) to 32 (worst) summarizing unpleasant affective states including fear, anxiety, embarrassment, hostility, stress, upset, irritation, and loneliness. The exposures are binary indicators of whether the participant communicated with at least one close contact via phone calls \(A_{\text{calls},t}\) or text messages \(A_{\text{texts},t}\). Close contacts are identified using a k-means clustering method (\cite{valeri2023}). Physical activity ($X_t$) has been demonstrated to be associated with negative mood as well as the tendency for social interaction and is thus included as a covariate (\cite{hamer2012}, \cite{peluso2005}). 

\cite{cai2025} analyzed the same BLS data using a state-space g-formula approach, focusing on subject-specific, time-varying effects of phone-based communication on mood over a 708-day follow-up, and reported a statistically significant contemporaneous protective effect of text messages on negative mood within a prespecified 81-day window for one participant, under the untestable assumption of no unmeasured confounding. However, in this observational setting it is plausible that important time-varying factors such as sleep quality and timing, adherence to psychotropic medications, and participation in psychotherapy sessions influence both text messaging behavior and daily mood yet are entirely unobserved, so that residual confounding may bias these individualized effect estimates. Thus, we would like to establish the robustness of a causal conclusion in their analysis. 

Following \cite{cai2025}, we use their reported time-varying effect analysis to identify the period during which the contemporaneous effect of text messages on negative mood was significant. Rather than searching for significance ourselves, we restrict attention to this pre-specified window and re-estimate the effect to examine whether the original finding remains robust in the presence of possible unmeasured confounding. 
Given the well-defined daily measures and variation in texting, the consistency and positivity assumptions are likely to be reasonable in this application. The strong temporal autocorrelation in mood and communication also supports the use of past outcomes and future treatments as relevant proxies. However, a mild violation of proxy independence is possible, because past negative mood $Y_{t-1}$ may affect current text-message exposure $A_{text,t}$. As shown in the simulation study, our estimator can still perform well under this type of mild violation.

We compare PCI with three linear models with different lags. The first linear model regresses the outcome on the current exposure and observed covariates only, the second adds first-order lagged outcomes, treatments, and covariates, and the third further includes second-order lagged terms. These models are simple baselines for assessing how much adjustment can be gained from including lagged history when proximal proxies are unavailable. In addition, we compare against a marginal structural mean model (MSMM) $\mathbb{E}[Y_t(a_t)]=\beta_0+\beta_1a_t$ using stabilized inverse-probability weights. For the MSMM, we estimate the contemporaneous effect by fitting two pooled models for the predicted probabilities across time, $\hat{p}_t^{den}=\mathbf{P}(A_t^{text}=1|A_{t-1}^{text}, X_t, Y_{t-1}, A_t^{call})$ and $\hat{p}_t^{num}=\mathbf{P}(A_t^{text}=1|A_{t-1}^{text})$. We then define per-time stablizied factor $sw_t=\frac{(\hat{p}_t^{num})^{A_t}(\hat{p}_t^{num})^{1-A_t}}{(\hat{p}_t^{den})^{A_t}(\hat{p}_t^{den})^{1-A_t}}$  with the cumulative weight $W_t=\prod_{k=1}^tsw_k$ used for the observation at time $t$ in the pooled analysis. Finally, we fit a weighted linear model with identity link and report the estimates and confidence intervals from a normal-approximation using the Newey–West robust standard error.

\subsection{Results}

Our data analysis focuses on a single participant and the prespecified window of days 561–641, during which prior work reported a significant contemporaneous effect of text messages on negative mood over a 708-day follow-up. In Table \ref{dataapptab}, baseline models that do not adjust for unmeasured confounding reproduce this finding, yielding statistically significant negative effects of text messages  at level 0.05, which is consistent with \cite{cai2025}. In contrast, the PCI estimators with $k=2, j=1$ and $k=3, j=2$ under the MSMM $g(a_{t}, V;\beta)=\beta_0+\beta_1a_t+\beta_2V$ produce an attenuated, statistically non-significant estimate closer to zero, indicating that adjusting for unmeasured confounding via proxies can materially alter inference on the contemporaneous text effect. We chose $k=2, j=1$ and $k=3, j=2$ because they performed well under the violation to proxy independence assumption in the simulation study that has a similar temporal structure as the application. Notably, the change in conclusions relative to Cai, Zeng, et al. (2024) does not arise simply because we adopt a different inferential framework. In this application, the simple linear models with lags and the marginal structural mean model (MSMM) yield very similar negative contemporaneous effects of text messaging on negative mood, whereas our proximal estimator produces an attenuated, non-significant effect that is closer to zero. This contrast reflects the fact that MSMMs and our PCI estimator both rely on weighting schemes but differ in how they adjust for time-varying unmeasured confounding through proxies, rather than a generic “method switch” per se. 

\begin{table}[H] 
\centering
\begin{tabular}{ccc}
        \toprule
         & \multicolumn{2}{c}{$A_{text}$} \\
        \cmidrule(r){2-3} 
         Estimators & $\hat{\beta}$ & (95\% CI)\\
        \midrule
        Linear model & -3.90 &(-5.48, -2.31) \\
        Linear model with lag-1 data & -1.46 & (-2.35, -0.56) \\
        Linear model with lag-2 data & -2.03 & (-3.45, -0.61) \\
        MSMM & -3.67 & (-5.32, -2.02) \\
        PCI ($k=2, j=1$) & 0.24 & (-3.39, 3.86) \\
        PCI ($k=3, j=2$) & 0.17 & (-7.63, 7.96) \\
        \bottomrule
    \end{tabular}

  \centering 
  \caption{Estimated contemporaneous effects of text messages on negative mood for a single participant in the Bipolar Longitudinal Study, using data from days 561--641. We report point estimates and 95\% confidence intervals (CIs) from the linear model, linear model with one- and two-step lags, marginal structural mean model (MSMM), and proximal causal inference (PCI) estimators with $(k=2,j=1)$ and $(k=3,j=2)$. CI: 95\% confidence interval, computed as point estimate $\pm\,1.96\times$ standard error under a normal approximation. Standard errors are model-based OLS standard errors for the linear models; Newey--West heteroskedasticity- and autocorrelation-consistent (HAC) standard errors (5 lags) for the MSMM; and standard errors derived from the estimator's asymptotic influence function (Equation~\ref{inffun}), which propagates uncertainty from the first-stage bridge-function estimation, for the PCI estimators. The PCI estimates target the contemporaneous effect under the proximal causal inference framework, while the linear and MSMM estimators serve as comparison methods.} 
  \label{dataapptab} 
\end{table}




\section{Discussion}

This paper develops a  proximal causal inference framework tailed to estimate the individual contemporaneous and lagged treatment effects in intensive time series and N-of-1 observational studies, where temporal ordering creates a natural source of proxy information through past outcomes and future treatments. A key contribution of this work is that it leverages variables that are already observed in longitudinal data streams, rather than requiring investigators to collect additional proxy variables by design. By doing so, the framework makes proximal adjustment more practical for mobile health and other person-level time series settings in which unmeasured confounding is a realistic concern. Our simulation study further shows how the method behaves under violations of identification assumptions that are unexplored in standard PCI settings, and highlights the value of incorporating lagged history when the data exhibit temporal dependence. In the Bipolar Longitudinal Study, we find that previously identified significant results were lost once incorporating the PCI framework to adjust for the unmeasured confounders.

An important strength of our PCI framework is that it converts the temporal structure of intensive longitudinal data into a source of identification, using past outcomes and future treatments as proxies for latent confounding. At the same time, the method inherits the usual limitations of proximal causal inference: completeness, proxy independence, and bridge-function assumptions are not directly testable, and the finite-sample behavior can be sensitive when these assumptions are only approximately satisfied. In addition, our estimation strategy relies on temporal regularity conditions such as stationarity or piecewise stable effects, which may not hold in every application. These limitations warrant careful scientific justification of the hypothesized time series structure. 

In practice, we recommend starting with small values of $k$(the temporal distance between proxies and the outcome and exposure) and $j$(the number of lags conditioned in the PCI framework), such as one or two lags, and then assessing sensitivity to larger values. Smaller $k$ places the proxy variables closer to the causal contrast of interest, which is often beneficial when temporal dependence is strong. Increasing $j$ can improve bias correction by incorporating richer recent history, but overly large $j$ may inflate variance, worsen numerical stability, and reduce the effective sample information. A reasonable strategy is therefore to select the smallest lag structure that is scientifically reasonable and then verify that conclusions are stable under nearby choices.

 While we illustrate the approach focusing on the contemporaneous treatment effect, results directly extend to the estimation of lagged treatment effects. An important future direction is to develop doubly robust estimators for contemporaneous effect as well as joint time-varying effects. Our estimator relies on correct specification of outcome confounding bridge functions. Doubly robust estimators can remain unbiased in large samples provided at least one confounding bridge function model is correct. Thus, it is important to develop alternative methods using treatment confounding bridge function as well (\cite{tchetgen2024}, \cite{ying2024}). 

\printbibliography 

\newpage

\section{Appendix}

\subsection{Assumptions for $q$-lagged effect}

\noindent \textbf{Assumption 1} Consistency:
\[Y_t=Y_t(A_{t-q}) \text{ almost surely}.\] 

\noindent \textbf{Assumption 2} Positivity:
\[0<Pr(A_{t-q}=a|A_{t-q-j:t-q-1}, U_{t-q-j:t-q}, L_{t-q-j:t-q})<1 \text{ for } a=1,0 \text{ almost surely}.\] 

\noindent \textbf{Assumption 3} Proxy independence:

\[Y_t\indep Z_{t-q} |A_{t-q}=a_{t-q},A_{t-q-j:t-q-1},X_{t-q-j:t-q},U_{t-q-j:t-q},\] \[W_t\indep (Z_{t-q},A_{t-q}) |X_{t-q-j:t-q},U_{t-q-j:t-q}.\] 

\noindent \textbf{Assumption 4} Bridge function:

There exists a function $h(w_{t-j:t},a_{t-q-j:t-q},x_{t-q-j:t-q})$ such that, \begin{align*}
    E[Y_t|Z_{t-q-j:t-q},A_{t-q-j:t-q}, X_{t-q-j:t-q}] = & \int h(w_{t-j:t},a_{t-q-j:t-q},x_{t-q-j:t-q})\\
    &dF(w_{t-j:t}|Z_{t-q-j:t-q},A_{t-q-j:t-q},X_{t-q-j:t-q}).
\end{align*}

\noindent \textbf{Assumption 5} Completeness:

(a)For any $a_{t-q-j:t-q}, x_{t-q-j:t-q}$, and any square-integrable function $v$,
\[E[v\{U_{t-q}\}|a_{t-q-j:t-q}, Z_{t-q-j:t-q}, x_{t-q-j:t-q}]=0 \text{ if and only if } v\{U_{t-q}\}=0 \text{ almost surely.}\]

(b)For any $a_{t-q-j:t-q}, x_{t-q-j:t-q}$, and any square-integrable function $v$,
\[E[v\{Z_{t-q}\}|a_{t-q-j:t-q}, W_{t-j:t}, x_{t-q-j:t-q}]=0 \text{ if and only if } v\{Z_{t-q}\}=0 \text{ almost surely.}\]
\subsection{Additional Simulation Studies}
\newpage
\begin{table}[H] 
\resizebox{\linewidth}{!}{%
\begin{tabular}{ccccccc}
        \toprule
        \multicolumn{7}{l}{\textbf{All assumptions are satisfied}} \\
        \midrule
         & \multicolumn{2}{c}{j=1} & \multicolumn{2}{c}{j=2} & \multicolumn{2}{c}{j=3} \\
        \cmidrule(r){2-3} \cmidrule(r){4-5} \cmidrule(r){6-7}
         k & $\hat{\beta}$ (95\% CI) & Coverage & $\hat{\beta}$ (95\% CI) & Coverage & $\hat{\beta}$ (95\% CI) & Coverage \\
        \midrule
        2 & -1.972 (-2.347, -1.597) & 94.2 & - &  & - & \\
        3 & -1.970 (-2.620, -1.320) & 97.3 & -1.946 (-7.063, 3.170) & 89.4 & - & \\
        4 & -1.955 (-2.757, -1.153) & 96.5 & -1.939 (-2.542, -1.336) & 92.3 & -1.950 (-3.046, -0.853) & 97.1 \\
    \end{tabular}
}
\resizebox{\linewidth}{!}{%
\begin{tabular}{ccccccc}
        \toprule
        \multicolumn{7}{l}{\textbf{Violation to completeness assumption}} \\
        \midrule
         & \multicolumn{2}{c}{j=1} & \multicolumn{2}{c}{j=2} & \multicolumn{2}{c}{j=3} \\
        \cmidrule(r){2-3} \cmidrule(r){4-5} \cmidrule(r){6-7}
         k & $\hat{\beta}$ (95\% CI) & Coverage & $\hat{\beta}$ (95\% CI) & Coverage & $\hat{\beta}$ (95\% CI) & Coverage \\
        \midrule
        2 & -2.047 (-2.932, -1.162) & 97.1 & - &  & - & \\
        3 & -2.138 (-3.335, -0.942) & 79.8 & -2.088 (-2.444, -1.732) & 72.7 & - & \\
        4 & -2.231 (-3.530, -0.934) & 76.6 & -2.127 (-3.025, -1.230) & 82.8 & -2.105 (-2.595, -1.615) & 77.2 \\
    \end{tabular}
}
\resizebox{\linewidth}{!}{%
\begin{tabular}{ccccccc}
        \toprule
        \multicolumn{7}{l}{\textbf{Violation to proxy independence assumption}} \\
        \midrule
         & \multicolumn{2}{c}{j=1} & \multicolumn{2}{c}{j=2} & \multicolumn{2}{c}{j=3} \\
        \cmidrule(r){2-3} \cmidrule(r){4-5} \cmidrule(r){6-7}
         k & $\hat{\beta}$ (95\% CI) & Coverage & $\hat{\beta}$ (95\% CI) & Coverage & $\hat{\beta}$ (95\% CI) & Coverage \\
        \midrule
        2 & -3.181 (-4.543, -1.819) & 10.3 & - &  & - & \\
        3 & -1.287 (-7.386, 4.812) & 63.7 & -2.582 (-4.294, -0.869) & 48.4 & - & \\
        4 & -1.329 (-2.363, -0.296) & 24.8 & -1.483 (-2.399, -0.567) & 30.9 & -2.126 (--31.673, 27.421) & 98.9  \\
    \end{tabular}
}
\resizebox{\linewidth}{!}{%
\begin{tabular}{ccccccc}
        \toprule
        \multicolumn{7}{l}{\textbf{Violation to proxy relevance assumption}} \\
        \midrule
         & \multicolumn{2}{c}{j=1} & \multicolumn{2}{c}{j=2} & \multicolumn{2}{c}{j=3} \\
        \cmidrule(r){2-3} \cmidrule(r){4-5} \cmidrule(r){6-7}
         k & $\hat{\beta}$ (95\% CI) & Coverage & $\hat{\beta}$ (95\% CI) & Coverage & $\hat{\beta}$ (95\% CI) & Coverage \\
        \midrule
        2 & -1.942 (-2.516, -1.368) & 74.2 & -  &  & - & \\
        3 & -1.927 (-2.486, -1.368) & 78.5 & -1.917 (-2.572, -1.262) & 72.5 & - & \\
        4 & -1.922 (-2.600, -1.245) & 82.4 & -1.917 (-2.362, -1.473) & 79.5 & -1.919 (-2.584, -1.254) & 92.1 \\
        \bottomrule
    \end{tabular}
}
  \centering 
  \caption{Simulation results for the contemporaneous effect under the main setting with follow-up length $T=10000$ and true effect $\beta=-2.0$. We report the average estimated effect $\hat{\beta}$, the average 95\% confidence interval, and empirical coverage for the \textbf{proximal causal inference (PCI) estimator} under the scenario where all assumptions are satisfied and under scenarios violating the completeness, proxy independence, or proxy relevance assumptions. The parameters $k,j$ determine the temporal placement of the proxy variables and the length of recent history included in the model, respectively.} 
  \label{tab5} 
\end{table}

\begin{table}[H] 
\resizebox{\linewidth}{!}{%
\begin{tabular}{ccccccc}
        \toprule
        \multicolumn{7}{l}{\textbf{All assumptions are satisfied}} \\
        \midrule
         & \multicolumn{2}{c}{lm} & \multicolumn{2}{c}{lm with lag-1 data} & \multicolumn{2}{c}{lm with lag-2 data} \\
        \cmidrule(r){2-3} \cmidrule(r){4-5} \cmidrule(r){6-7}
         k & $\hat{\beta}$ (95\% CI) & Coverage & $\hat{\beta}$ (95\% CI) & Coverage & $\hat{\beta}$ (95\% CI) & Coverage \\
        \midrule

        2 & -2.203 (-2.258, -2.148) & 0.0 & -1.947 (-1.975, -1.919) & 4.1 & -1.947 (-1.974, -1.919) & 3.7 \\
    \end{tabular}
}
\resizebox{\linewidth}{!}{%
\begin{tabular}{ccccccc}
        \toprule
        \multicolumn{7}{l}{\textbf{Violation to completeness assumption}} \\
        \midrule
         & \multicolumn{2}{c}{lm} & \multicolumn{2}{c}{lm with lag-1 data} & \multicolumn{2}{c}{lm with lag-2 data} \\
        \cmidrule(r){2-3} \cmidrule(r){4-5} \cmidrule(r){6-7}
         k & $\hat{\beta}$ (95\% CI) & Coverage & $\hat{\beta}$ (95\% CI) & Coverage & $\hat{\beta}$ (95\% CI) & Coverage \\
        \midrule

        2 & -1.065 (-1.139, -0.991) & 0.0 & -1.814 (-1.849, -1.780) & 0.0 & -1.855 (-1.889, -1.822) & 0.0 \\
    \end{tabular}
}
\resizebox{\linewidth}{!}{%
\begin{tabular}{ccccccc}
        \toprule
        \multicolumn{7}{l}{\textbf{Violation to proxy independence assumption}} \\
        \midrule
         & \multicolumn{2}{c}{lm} & \multicolumn{2}{c}{lm with lag-1 data} & \multicolumn{2}{c}{lm with lag-2 data} \\
        \cmidrule(r){2-3} \cmidrule(r){4-5} \cmidrule(r){6-7}
         k & $\hat{\beta}$ (95\% CI) & Coverage & $\hat{\beta}$ (95\% CI) & Coverage & $\hat{\beta}$ (95\% CI) & Coverage  \\
        \midrule
        2 & -0.908 (-0.998, -0.819) & 0.0 & -1.446 (-1.521, -1.371) & 0.0 & -1.452 (-1.527, -1.378) & 0.0 \\
    \end{tabular}
}
\resizebox{\linewidth}{!}{%
\begin{tabular}{ccccccc}
        \toprule
        \multicolumn{7}{l}{\textbf{Violation to proxy relevance assumption}} \\
        \midrule
         & \multicolumn{2}{c}{lm} & \multicolumn{2}{c}{lm with lag-1 data} & \multicolumn{2}{c}{lm with lag-2 data} \\
        \cmidrule(r){2-3} \cmidrule(r){4-5} \cmidrule(r){6-7}
         k & $\hat{\beta}$ (95\% CI) & Coverage & $\hat{\beta}$ (95\% CI) & Coverage & $\hat{\beta}$ (95\% CI) & Coverage\\
        \midrule
        2 & -2.184 (-2.238, -2.130) & 0.0 & -1.921 (-1.950, -1.893) & 0.0 & -1.922 (-1.950, -1.893) & 0.0 \\
        \bottomrule
    \end{tabular}
}
  \centering 
  \caption{Simulation results for the contemporaneous effect under the main setting with follow-up length $T=10000$ and true effect $\beta=-2.0$, based on \textbf{linear model comparison methods}. We report the average estimated effect $\hat{\beta}$, the average 95\% confidence interval, and empirical coverage for models with no lagged history, with one lag, with two lags. Results are shown under the scenarios where the completeness, proxy independence, or proxy relevance assumptions are satisfied or violated.} 
  \label{tab6} 
\end{table}

\begin{table}[H] 
\resizebox{\linewidth}{!}{%
\begin{tabular}{ccccccc}
        \toprule
        \multicolumn{7}{l}{\textbf{When all assumptions are satisfied}} \\
        \midrule
         & \multicolumn{2}{c}{j=1} & \multicolumn{2}{c}{j=2} & \multicolumn{2}{c}{j=3} \\
        \cmidrule(r){2-3} \cmidrule(r){4-5} \cmidrule(r){6-7}
         k & $\hat{\beta}$ (95\% CI) & Coverage & $\hat{\beta}$ (95\% CI) & Coverage & $\hat{\beta}$ (95\% CI) & Coverage \\
        \midrule
        3 & -1.167 (-3.577, 1.243) & 99.0 & - &  & - & \\
        4 & -1.112 (-1.874, -0.350) & 89.1 & -1.114 (-1.970, -0.259) & 97.0 & - &  \\
        5 & -1.106 (-1.599, -0.613) & 86.4 & -1.113 (-1.477, -0.748) & 94.9 & -1.102 (-1.634, -0.571) & 96.2  \\
        \bottomrule
    \end{tabular}
}
  \centering 
  \caption{Simulation results for  the lagged effect under the main setting with follow-up length $T=10000$ and true effect $\beta=-1.154$. We provide the estimated $\beta$, average 95\% confidence intervals, and coverages from linear models with lagged data.} 
  \label{tab_lag7} 
\end{table}

\begin{table}[H] 
\resizebox{\linewidth}{!}{%
\begin{tabular}{ccccccc}
        \toprule
        \multicolumn{7}{l}{\textbf{All assumptions are satisfied T=10000}} \\
        \midrule
         & \multicolumn{2}{c}{lm} & \multicolumn{2}{c}{lm with lag-1 data} & \multicolumn{2}{c}{lm with lag-2 data} \\
        \cmidrule(r){2-3} \cmidrule(r){4-5} \cmidrule(r){6-7}
         k & $\hat{\beta}$ (95\% CI) & Coverage & $\hat{\beta}$ (95\% CI) & Coverage & $\hat{\beta}$ (95\% CI) & Coverage \\
        \midrule

        3 & -1.193 (-1.225, -1.160) & 35.5 & -1.103 (-1.133, -1.072) & 8.4 & -1.105 (-1.135, -1.075) & 10.1 \\
    \end{tabular}
}
\resizebox{\linewidth}{!}{%
\begin{tabular}{ccccccc}
        \toprule
        \multicolumn{7}{l}{\textbf{All assumptions are satisfied T=90}} \\
        \midrule
         & \multicolumn{2}{c}{lm} & \multicolumn{2}{c}{lm with lag-1 data} & \multicolumn{2}{c}{lm with lag-2 data} \\
        \cmidrule(r){2-3} \cmidrule(r){4-5} \cmidrule(r){6-7}
         k & $\hat{\beta}$ (95\% CI) & Coverage & $\hat{\beta}$ (95\% CI) & Coverage & $\hat{\beta}$ (95\% CI) & Coverage \\
        \midrule

        3 & -1.163 (-1.513, -0.813) & 92.5 & -1.081 (-1.419, -0.744) & 92.8 & -1.089 (-1.425, -0.752) & 93.9 \\
        \bottomrule
    \end{tabular}
}
  \centering 
  \caption{Simulation results for the lagged effect under the setting with follow-up length $T=10000$ and true effect $\beta=-1.154$, based on \textbf{linear model comparison methods}. We report the average estimated effect $\hat{\beta}$, the average 95\% confidence interval, and empirical coverage for models with no lagged history, with one lag, with two lags. Results are shown under the scenarios where the completeness, proxy independence, or proxy relevance assumptions are satisfied or violated.} 
  \label{tab8} 
\end{table}

\end{document}